\UseRawInputEncoding  

\documentclass[times,twocolumn,final]{elsarticle}

\usepackage{lineno}
\usepackage{framed,multirow}
\usepackage{amssymb}
\usepackage{latexsym}
\usepackage{bbm}

\usepackage{url}
\usepackage{xcolor}

\usepackage{graphicx}
\usepackage{amsmath}
\usepackage{amssymb}
\usepackage{verbatim}
\usepackage{booktabs}
\usepackage{array}
\usepackage{epstopdf}
\usepackage{makecell}
\usepackage{tabularx} 
\usepackage{booktabs}   
\usepackage{mathrsfs} 
\usepackage{framed, multirow} 
\usepackage{amsfonts} 
\usepackage{wrapfig}  
\usepackage{cite}
\usepackage{footnote}

\usepackage{latexsym}
\usepackage[caption = false]{subfig}
\usepackage{graphicx}

\usepackage{tablefootnote}
\usepackage{float}
\usepackage[hidelinks,colorlinks=true,linkcolor=blue,citecolor=blue]{hyperref}

\def\x{{\bf x}}

\def\0{{\bf 0}}
\def\1{{\bf 1}}

\usepackage{geometry}
\begin{document}
	
	\begin{frontmatter}
		\title{Leveraging Task-Specific Knowledge from LLM for Semi-Supervised 3D Medical Image Segmentation}

\author[1]{Suruchi Kumari}\ead{suruchi\_k@cs.iitr.ac.in}
\author[2]{Aryan Das} \ead{aryan.das2021@vitbhopal.ac.in}
\author[3]{Swalpa Kumar Roy} \ead{swalpa@agemc.ac.in}
\author[4]{Indu Joshi} \ead{indujoshi@iitmandi.ac.in}

\author[1]{Pravendra~Singh\corref{mycorrespondingauthor}}\ead{pravendra.singh@cs.iitr.ac.in}

\cortext[mycorrespondingauthor]{Corresponding author: Pravendra Singh }

\address[1]{Department of Computer Science and Engineering, Indian Institute of Technology Roorkee, India}
\address[2]{School Of Computer Science Engineering, Vellore Institute of Technology, Bhopal, Madhya Pradesh, India}

\address[3]{Alipurduar Government Engineering and Management College, India}
\address[4]{Indian Institute of Technology Mandi, India}

\begin{abstract}
Traditional supervised 3D medical image segmentation models need voxel-level annotations, which require huge human effort, time, and cost. Semi-supervised learning (SSL) addresses this limitation of supervised learning by facilitating learning with a limited annotated and larger amount of unannotated training samples. However, state-of-the-art SSL models still struggle to fully exploit the potential of learning from unannotated samples. To facilitate effective learning from unannotated data, we introduce \textit{LLM-SegNet}, which exploits a \textit{large language model (LLM)} to integrate task-specific knowledge into our co-training framework. This knowledge aids the model in comprehensively understanding the features of the region of interest (ROI), ultimately leading to more efficient segmentation. Additionally, to further reduce erroneous segmentation, we propose a \textit{Unified Segmentation} loss function. This loss function reduces erroneous segmentation by not only prioritizing regions where the model is confident in predicting between foreground or background pixels but also effectively addressing areas where the model lacks high confidence in predictions. Experiments on publicly available Left Atrium, Pancreas-CT, and Brats-19 datasets demonstrate the superior performance of LLM-SegNet compared to the state-of-the-art. Furthermore, we conducted several ablation studies to demonstrate the effectiveness of various modules and loss functions leveraged by LLM-SegNet. 
\end{abstract}

\begin{keyword}
3D medical image segmentation \sep Semi-supervised learning \sep Deep learning\sep Large language model
\end{keyword}

\end{frontmatter}


\section{Introduction \label{Intro}}

Training a convolutional neural network (CNN) to segment 3D medical images is a challenging problem. Traditionally, this problem has been addressed as a supervised learning problem that requires voxel-level manually annotated samples. However, the cost, domain expertise, and human effort required to collect such annotations create a need to develop techniques that facilitate learning with fewer annotated samples \citep{chen2022recent, kumari2023data}. Semi-supervised learning is one of the most promising directions in which few annotated, and a large number of unannotated data are used to train the CNN to segment 3D medical images.

Among the literature spanning semi-supervised learning (described in detail in Section \ref{RL_SSL} ), due to the ease of implementation, generalization ability, and flexibility that \textit{co-training} offers, it turns to be a leading method for semi-supervised learning. To be specific, for each data point in the dataset, the co-training framework \citep{qiao2018deep} assumes the existence of two distinct and complementary views, each capable of training a segmentation model effectively. A separate segmentation model is trained using each view. Co-training relies on the assumption that the segmentation models trained on the first and second views should yield consistent predictions for the dataset. In principle, the success of co-training relies on the distinctiveness and complementarity of the two views \citep{yang2022survey, kumari2023data}. However, due to the challenges involved in learning from unlabeled samples, the performance of the state-of-the-art semi-supervised 3D medical image segmentation techniques is still far from satisfactory.

On the other hand, the field of \textit{large language models (LLMs)} has made significant advancements in recent years. Unlike other AI models, LLMs have the ability to comprehend and produce natural language, as well as perform language-based reasoning. This enables them to adapt to various tasks without requiring specific training for each one. Consequently, LLMs have proven to be effective in giving medical expertise and guidance. ChatGPT \citep{chatgpt2023optimizing}, a big dialog-based LLM, has proven to be highly effective in solving several complex tasks. Various studies have employed LLMs for diverse medical imaging tasks \citep{wang2023chatcad, chen2023generative, lee2023llm}. However, all these methods utilize Vision-Language Pretraining (VLP) to integrate information from both images and text. Furthermore, most of the methods leveraging LLMs explore VLP for 2D images, but 3D images remain largely unexplored due to the scarcity of associated 3D medical image-text datasets.

\par Obtaining paired 3D medical image-text datasets from LLM is very time-consuming as generating corresponding text associated with the image takes a lot of time. Another major shortcoming that hinders the application of LLMs in 3D images is that, till now, LLM only accepts 2D images as input. In order to exploit existing LLMs for 3D images, a 3D image must be split into multiple 2D images, which can then be given as input to LLM. Subsequently, the LLM will be required to generate associated text information for each 2D image. As a result, such a step can significantly increase processing time, both at the training and testing time, as well as, make the 3D medical image segmentation heavily reliant on LLM. To address this limitation of LLMs for 3D images while leveraging its usefulness in compensating for limited annotations, we introduce LLM-SegNet, which aims to explore the use of LLM to generate task-specific knowledge (see Figure \ref{fig:prompt}) and use this knowledge to improve semi-supervised 3D medical image segmentation. To the best of our knowledge, LLM-SegNet is the first work to utilize task-specific information from LLM in a simple yet highly effective co-training framework, which uses CNN as the backbone for semi-supervised 3D medical image segmentation. Additionally, since we are not providing images as input to the LLM but instead generating task-specific knowledge directly from the LLM, our approach is highly efficient at both training and testing times and is not bottlenecked by LLM inference time. Moreover, the task-specific knowledge generated from the LLM can be used both while training and testing LLM-SegNet, facilitating improved 3D segmentation performance.

Semi-supervised learning leverages not only annotated data but also unannotated data. Unsupervised loss plays a key role in learning from unannotated data. Considering this aspect, rather than utilizing conventional mean squared error (MSE) or Negative Log Likelihood (NLL) loss functions, we introduce \textit{unified segmentation loss}. This loss function not only focuses on regions where the model is confident but also addresses areas where it lacks high confidence in foreground and background predictions. Specifically, the NLL loss encourages the model to make confident and discriminative predictions by penalizing incorrect classifications more heavily. In contrast, MSE loss measures the squared difference between predictions and pseudo-labels, which tends to be less sensitive to label noise compared to NLL loss. Motivated by this observation, we apply NLL loss when our model makes confident predictions about the foreground and background. However, when the model lacks high confidence in its foreground and background predictions, we opt for mean squared error loss.

To summarize, we tackle the challenges of learning from unannotated data by introducing a co-training approach. In this method, we leverage large language models to integrate task-specific knowledge into our co-training framework. This knowledge can assist the model in effectively understanding the characteristics of the region of interest (ROI), ultimately facilitating more efficient segmentation.  Additionally, we introduce a unified segmentation loss function that not only focuses on regions where the model is confident but also addresses areas where it lacks high confidence in the foreground and background predictions. Results on publicly available the Left Atrium (LA) \citep{XiongXHHBZVRMYH21}, Pancreas-CT (Pancreas) \citep{ClarkVSFKKMPMPTP13}, and Brats19 \citep{bakas2020brats} datasets demonstrate the superior segmentation ability of LLM-SegNet compared to the state-of-the-art.

\section{Related work \label{sec_background}}

\subsection{Semi-supervised medical image segmentation \label{RL_SSL}}

\begin{figure}
    \centering
    \includegraphics[width=0.42\textwidth]{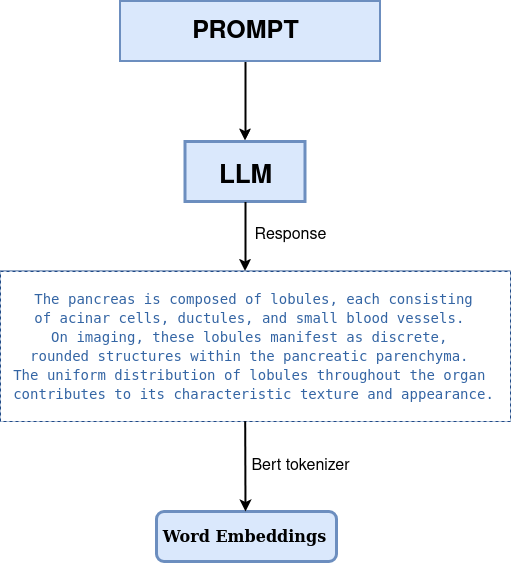}
    \caption{Illustration of the process for creating specific word embeddings. Initially, we provide prompt input to LLM, which generates a textual response. This response is then fed into the BERT tokenizer to produce the corresponding word embeddings.}
    \label{fig:prompt}
\end{figure}

Existing semi-supervised medical image segmentation approaches can be roughly categorized into \textit{adversarial learning-based methods} \citep{nguyen2023cross, lee2022voxel}, \textit{pseudo-labeling-based methods} \citep{zhang2024quality, kumari2024leveraging}, and \textit{consistency regularization based methods} \citep{you2023action++, gai2024sdmi, lu2024dual, mahmood2024splal}.

Methods leveraging adversarial learning for semi-supervised learning (SSL) exploit a discriminator network to ensure that the output of the segmentation model (generator) is indistinguishable from the ground truth annotations. For instance, Nguyen-Duc \textit{et al.} propose a cross-adversarial local distribution regularization \citep{nguyen2023cross} that ensures that the output of similar but perturbed samples is similar. Wu \textit{et al.} propose SS-Net \citep{wu2022exploring} that ensures smoothness over adversarial perturbations while enabling higher inter-class separation. 
However, adversarial learning-based methods suffer from increased complexity primarily due to adversary training \citep{pelaez2023survey}. This added complexity can complicate implementation and require more computational resources.

In SSL, consistency regularization methods aim to maintain the stability of predictions for unlabeled examples, even when subjected to different perturbations such as noise or data augmentation. By ensuring that predictions remain consistent across different perspectives or alterations of the same unlabeled instance, these methods enable the model to learn more robust and generalizable representations from the unlabeled data. One notable approach for consistency regularization is self-ensembling Mean Teacher (MT), which applies supervised loss on labeled data in addition to perturbation-based consistency loss between the self-ensembling teacher model and the student model on unlabeled data. Some of the methods utilizing the MT framework include Yu \textit{et al.} \citep{yu2019uncertainty} and You \textit{et al.} \citep{you2023action++} that use the MT framework for left atrium segmentation. Similarly, Bai \textit{et al.} \citep{bai2023bidirectional} employ a teacher-student architecture for 3D medical image segmentation. They propose a method where annotated foregrounds are copied and pasted onto unannotated backgrounds and vice versa. These mixed images are then forwarded to the student model. The student model is trained using annotations from the labeled samples and pseudo-annotations obtained from the teacher model. However, consistency regularization methods may struggle with data distributions that deviate significantly from the assumptions made during training, leading to suboptimal performance in real-world scenarios.

Lastly, in SSL, pseudo-labeling methods involve assigning pseudo-labels to unannotated samples using guidance from annotated examples. Pseudo-labeling methods can be broadly categorized into \textit{self-training} and \textit{co-training}. In self-training, labeled data is initially used to train a model, which then generates predictions for unlabeled data. These unlabeled predictions are subsequently utilized to further train the model using the unlabeled data. On the other hand, in co-training, multiple models are employed, all trained on labeled data. However, for unlabeled data, they rely on predictions from each other. Several approaches have been proposed to utilize a pseudo-labeling framework for medical image segmentation. For instance, wang \textit{et al.} proposed a mutual correction framework \citep{Wang0B0023} that addresses semi-supervised learning as a network bias correction problem to guide both networks in co-training. Xie \textit{et al.} propose deep mutual distillation \citep{xie2023deep} with temperature scaling that enables reliable mutual guidance in uncertain regions.
Miao \textit{et al.} \citep{miao2023caussl} propose CauSSL, a causality-based framework that improves the algorithmic independence of the co-training framework. Wang and Li propose a dual-debiased heterogeneous co-training framework \citep{wang2023dhc} to mitigate the issue of class imbalance in semi-supervised learning. Gao \textit{et al.} propose a correlation-aware mutual learning framework \citep{gao2023correlation} that establishes cross-sample correlations to enable knowledge transfer between annotated and unannotated data. Although the methods mentioned above have achieved great success, they still struggle to effectively learn from unlabeled data.
Self-training iteratively labels unlabeled data based on its own predictions, potentially reinforcing biases and propagating errors, especially with challenging data. In contrast, co-training leverages diverse data perspectives to mitigate the impact of noisy labels, resulting in more robust models, particularly with uncertain data. Therefore, we adopt the co-training framework for our proposed method.

\begin{figure*}
    \centering
    \includegraphics[width=0.8\textwidth]{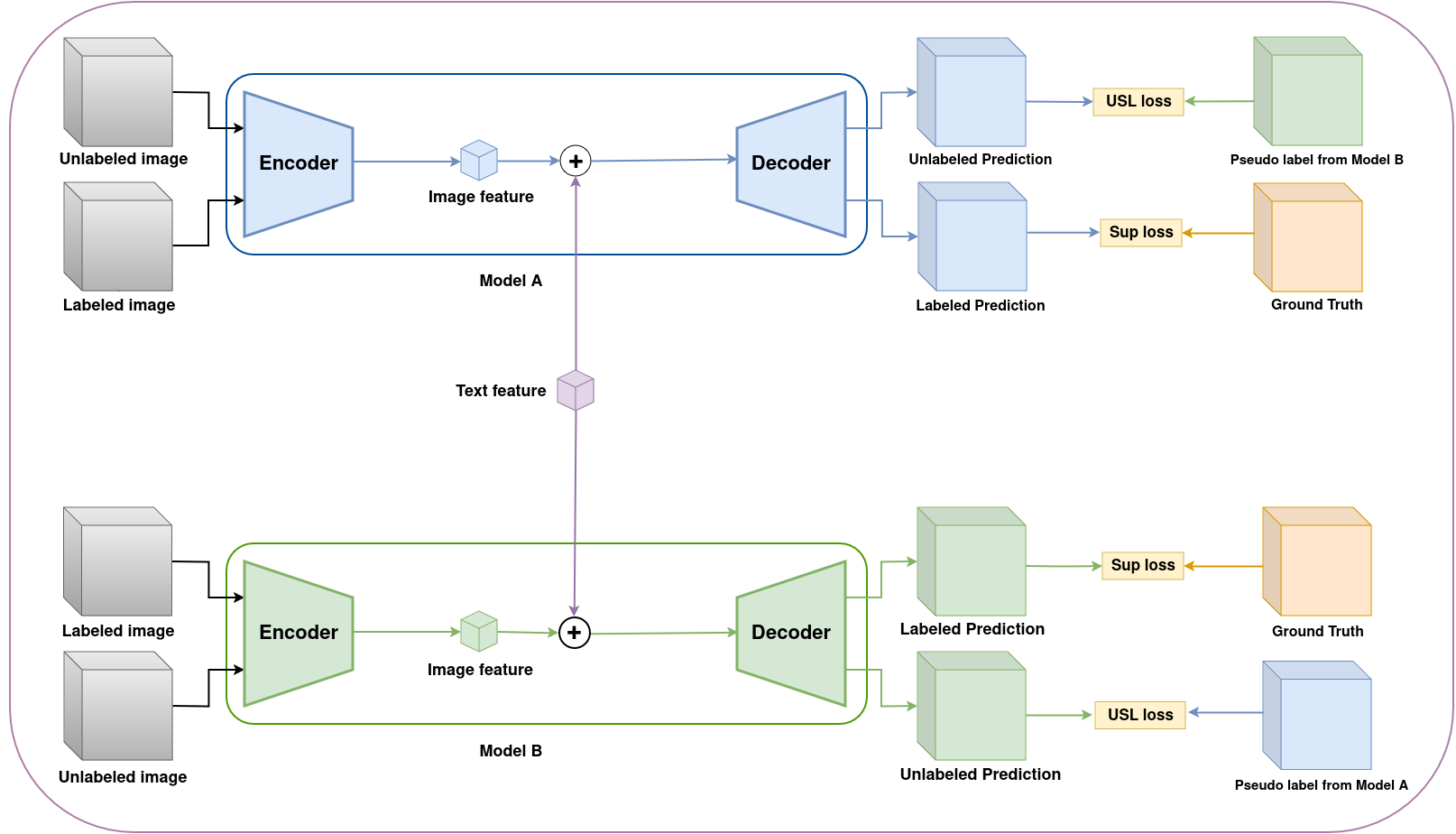}
    \caption{Overview of our LLM-SegNet approach. We apply the supervised loss to align the labeled predictions from model $\mathcal{A}$ with the ground truth and do the same for model $\mathcal{B}$. To learn from unlabeled data, we utilize the proposed unified segmentation loss (USL) between the pseudo-label from the other model and the model's own unlabeled predictions. Furthermore, our co-training framework uses both text and image features jointly to train the model.}
    \label{fig:main}
\end{figure*}

\begin{table*}[t]
\centering
\caption{Sample responses obtained from LLM for the following prompt for LA dataset: \textit{Generate 15 descriptive crucial features, each of a minimum of 60 words for effectively segmenting the left atrium from 3D MRI images, offering detailed insights into organ shape, structure, foreground and background details, anatomical landmarks, intensity characteristics, surrounding structures, contextual information. Consider including structural features such as the crescent shape, smooth endocardial surface, pulmonary vein ostia, atrial appendage, fossa ovalis, mitral valve annulus, left auricle, atrial septum, smooth muscle bundles, and papillary muscles. Each description should be distinctive to provide comprehensive guidance for successful segmentation.} Similarly, sample responses obtained from LLM for the following prompt for the pancreas dataset: \textit{Generate 15 descriptive crucial features, each of a minimum of 60 words for effectively segmenting the pancreas from 3D CT scan images, offering detailed insights into organ shape, structure, foreground and background details, anatomical landmarks, intensity characteristics, surrounding structures, and contextual information. Consider including structural features such as the pancreatic duct, islets of Langerhans, acinar cells, lobules, head, body, tail, surrounding vessels, and surrounding organs. Each description should be distinctive to provide comprehensive guidance for successful segmentation.
} Res refers to Response.}
\label{tab:LLM}

\begin{tabular}{|p{2cm}|p{2cm}|p{12cm}|}
\hline
\textbf{Dataset}& \textbf{Response} & \textbf{Output of LLM}  \\ \hline
\cline{2-3}  \multirow{3}{*}{LA} &  Res1 & The left atrium typically exhibits a crescent shape with a convex lateral wall and a concave medial wall. Its smooth endocardial surface lacks prominent protrusions, facilitating segmentation by distinguishing it from adjacent structures. \\ \hline
 LA &   Res2& Distinct pulmonary vein ostia are discernible along the posterior aspect of the left atrium, serving as crucial landmarks for segmentation. These openings mark the entry points of the pulmonary veins into the atrium and can be identified based on their location and contrast intensity.\\ \hline
LA  & Res3 & Papillary muscles are prominent structures located within the left ventricle, but their proximity to the left atrium can influence segmentation. Their distinctive appearance and position relative to the atrioventricular valves serve as important references for delineating the left atrium.\\ \hline
Pancreas & Res1 & Internally, the pancreas comprises lobules organized into lobes, with a dense network of interconnected acinar cells and ducts. The lobules are discernible as small, rounded clusters of cells. The pancreatic duct, which traverses through the organ, is a prominent linear feature, typically running longitudinally from the tail towards the head.\\ \hline
Pancreas & Res2 & The pancreas is closely associated with several neighboring structures, including the duodenum, spleen, stomach, liver, superior mesenteric vessels, and splenic vessels. The duodenum, in particular, wraps around the head of the pancreas, forming the C-loop, while the splenic artery and vein course along the superior border of the pancreas. \\ \hline

\end{tabular}
\label{tab:llm}
\end{table*}

\subsection{LLM}

The field of large language models (LLMs) has made significant advancements in recent years. Unlike other AI models, LLMs have the ability to comprehend and produce natural language, as well as perform language-based reasoning. This enables them to adapt to various tasks without requiring specific training for each one. GPT-3 \citep{brown2020language}, an autoregressive language model with 175 billion parameters, has demonstrated strong performance across numerous natural language processing tasks. Several LLMs undergo training tailored to specific tasks. For instance, models like BioBERT \citep{lee2020biobert} and ClinicalBERT is trained over the BERT model using PubMed data, specifically focusing on biomedical information. LLMs have been used in several studies for a variety of medical imaging applications. For example, Wang et al. \citep{wang2023chatcad} use LLMs for interactive computer-aided diagnosis on medical images. Moreover, Chen et al. \citep{chen2023generative} use LLMs in the field of 3D medical image segmentation to produce medical-style text from 3D medical images, then use this artificial text to oversee the learning of 3D visual representations. Lee et al. \citep{lee2023llm} introduce LLM-CXR, a bidirectional reasoning generation mechanism. This approach encodes images into tokens, enabling LLMs to process and generate both text and images. However, all these methods utilize Vision-Language pretraining to merge information from both images and text, yet there has been no attempt to integrate information from LLMs into a CNN architecture.

\subsection{Vision language pre-training}

Vision-language models (VLMs) integrate information from both images and text, enabling them to comprehend and generate content that combines visual and linguistic cues. By leveraging large-scale pre-training on vast multimodal datasets, VLMs can learn rich representations of the relationships between images and associated text, facilitating a wide range of downstream tasks such as image captioning, visual question answering, and image-text retrieval. Some of the current representative VLM models include CLIP \citep{radford2021learning} and ALIGN \citep{jia2021scaling}. These models align image representations with corresponding text representations directly through a contrastive loss. While other methods utilize attention mechanisms to incorporate multi-modal information. For example, BLIP \citep{li2022blip} and CoCa \citep{yu2022coca} integrate a decoder and include image-to-text generation as an additional task. Flamingo \citep{alayrac2022flamingo} utilizes gated cross-attention to combine language and visual modalities after freezing a pre-trained vision encoder and language model.

For label-efficient multimodal medical imaging representation, GLoRIA \citep{huang2021gloria} utilizes an attention-based framework to learn both global and local representations between medical reports and images. Inspired by CLIP \citep{radford2021learning}, wang et al. \citep{wang2022medclip} propose MedCLIP, which decouples paired images and texts and employs soft targets representing semantic similarities to train on unpaired medical images and texts. CXR-CLIP \citep{you2023cxr} addresses the lack of image-text data in chest X-rays by converting image-label pairs into image-text pairs through a generic prompt and utilizing multiple images and sections in a radiologic report. Despite its significance, the lack of publicly available datasets with 3D medical image-text pairs limits VLM's applicability to a larger range of medical applications.



\section{Method}
Mathematically, we define the 3D volume of a medical image as ${X} \in \mathbb{R}^{W \times H \times L}$. The objective of semi-supervised medical image segmentation is to predict the per-voxel label map $\widehat{{Y}} \in \{0, 1, \ldots, K-1\}^{W \times H \times L}$, indicating the locations of the background and foreground region of interest within $\mathbf{X}$, where $K$ represents the number of classes. Let $X_{ann}$ signify the distribution of 3D labeled medical images of interest and the corresponding voxel-level annotations, i.e.,  $X_{ann}=\{(x, y)\}$. $X_{unann}$ denotes the distribution of unannotated 3D medical images of interest, i.e., $X_{unann}=\{x\}$. For semi-supervised learning, we utilize $X_{ann}$ and $X_{unann}$ to train a model $f(.)$ to generate segmentation labels.

\begin{table*}[t] \small 
\centering
\scriptsize
\setlength{\tabcolsep}{0.8mm}{
\resizebox{0.8\textwidth}{!}{
\begin{tabular}{c|cc|llll}
\hline
\multicolumn{1}{c|}{\multirow{2}{*}{Method}} & \multicolumn{2}{c|}{Scans used}  & \multicolumn{4}{c}{Metrics} \\ 
\cline{2-7} \multicolumn{1}{c|}{} & \multicolumn{1}{l}{Annotated} & \multicolumn{1}{l|}{Unannotated} & 
Dice$\uparrow$ & Jaccard$\uparrow$ & 95HD$\downarrow$ & ASD$\downarrow$ \\ \hline
\multicolumn{1}{c|}{V-Net \citep{MilletariNA16}} &\multicolumn{1}{c}{4(5\%)} &\multicolumn{1}{c|}{0} &52.55 &39.60 &47.05 &9.87 \\ 
\multicolumn{1}{c|}{V-Net \citep{MilletariNA16}} &\multicolumn{1}{c}{8(10\%)} &\multicolumn{1}{c|}{0} &82.74 &71.72 &13.35 &3.26 \\ \hline
UA-MT \citep{yu2019uncertainty} (MICCAI’19) &  & 
 & 82.26 & 70.98 & 13.71 & 3.82 \\
SASSNet \citep{li2020shape} (MICCAI'20)&  \multirow{7}{*}{4(5\%)} & \multirow{7}{*}{76(95\%)}   & 81.60 & 69.63 & 16.16 & 3.58 \\ 
DTC \citep{luo2021semi} (AAAI'21) & & & 81.25 & 69.33 & 14.90 & 3.99 \\
URPC \citep{luo2021efficient} (MICCAI'21) & & & 82.48 & 71.35 & 14.65 & 3.65 \\ 
SS-Net \citep{wu2022exploring} (MICCAI'22) & & & 86.33 & 76.15 & 9.97 & 2.31 \\ 
ACTION++ \citep{you2023action++} (MICCAI'23) & & & 87.8 & NA & NA & 2.09   \\ 
BCP \citep{bai2023bidirectional} (CVPR'23) & & & 88.02 & 78.72 & 7.90 & 2.15 \\ 
Ours & & & \textbf{88.63}{\color{red}\textbf{\scriptsize{$\uparrow$}0.61}}&\textbf{79.53}{\color{red}\textbf{\scriptsize{$\uparrow$}0.81}}&\textbf{7.62}{\color{red}\textbf{\scriptsize{$\downarrow$}0.28}}&\textbf{2.07}{\color{red}\textbf{\scriptsize{$\downarrow$}0.02}} \\                 \hline

 UA-MT \citep{yu2019uncertainty} (MICCAI’19) & \multirow{7}{*}{8(10\%)} & \multirow{7}{*}{72(90\%)} 
& 87.79 & 78.39 & 8.68 & 2.12 \\
SASSNet \citep{li2020shape} (MICCAI'20) &  &   & 87.54 & 78.05 & 9.84 & 2.59 \\ 
DTC \citep{luo2021semi} (AAAI'21) & & & 87.51 & 78.17 & 8.23 & 2.36  \\
URPC \citep{luo2021efficient} (MICCAI'21) & & & 86.92 & 77.03 & 11.13 & 2.28  \\ 
SS-Net \citep{wu2022exploring} (MICCAI'22) & & & 88.55 & 79.62 & 7.49 & 1.90 \\

BCP \citep{bai2023bidirectional} (CVPR'23) & & & 89.62 & 81.31 & 6.81 & 1.76 \\
ACTION++ \citep{you2023action++} (MICCAI'23) & & & 89.9 & NA & NA & 1.74   \\ 
DMD \citep{xie2023deep} (MICCAI'23) & & & 89.70 &  81.42 &  6.88 & 1.78 \\
MLRPL \citep{su2024mutual} (MIA' 2024)  & & & 89.86 & 81.68 & 6.91 & 1.85 \\
Ours & & & \textbf{91.45}{$\color{red}\textbf{\scriptsize{$\uparrow$}1.59}$}&\textbf{84.31}{\color{red}\textbf{\scriptsize{$\uparrow$}2.63}}&\textbf{4.66}{\color{red}\textbf{\scriptsize{$\downarrow$}2.15}}&\textbf{1.62}{\color{red}\textbf{\scriptsize{$\downarrow$}0.12}}\\ \hline
\end{tabular}}
}

\caption{Comparisons with state-of-the-art semi-supervised segmentation methods on LA dataset. 
Improvements compared with the second-best results are highlighted in \textcolor{red}{Red}.}
\label{tab:LA8}
\end{table*}

\begin{table*}[t] 
\centering
\scriptsize
\setlength{\tabcolsep}{1mm}{
\resizebox{0.8\textwidth}{!}{
\begin{tabular}{c|cc|llll}
\hline
\multicolumn{1}{c|}{\multirow{2}{*}{Method}} & \multicolumn{2}{c|}{Scans used}  & \multicolumn{4}{c}{Metrics} \\ 
\cline{2-7} \multicolumn{1}{c|}{} & \multicolumn{1}{l}{Annotated} & \multicolumn{1}{l|}{Unannotated} & 
Dice$\uparrow$ & Jaccard$\uparrow$ & 95HD$\downarrow$ & ASD$\downarrow$ \\ \hline
\multicolumn{1}{c|}{V-Net} &\multicolumn{1}{c}{6(10\%)} &\multicolumn{1}{c|}{0} & 54.94 & 40.87 & 47.48 & 17.43 \\ 
\multicolumn{1}{c|}{V-Net} &\multicolumn{1}{c}{12(20\%)} &\multicolumn{1}{c|}{0} & 71.52 & 57.68 & 18.12 & 5.41 \\ \hline

 UA-MT \citep{yu2019uncertainty} (MICCAI’19) & \multirow{7}{*}{6(10\%)} & \multirow{7}{*}{56(90\%)} & 66.44 & 52.02 & 17.04 & 3.03 \\
 SASSNet \citep{li2020shape} (MICCAI'20) &  &  & 68.97 & 54.29 & 18.83 & 1.96   \\ 
DTC \citep{luo2021semi} (AAAI'21) &  &  & 66.58 & 51.79 & 15.46 & 4.16 \\
URPC \citep{luo2021efficient} (MICCAI'21) & & & 73.53 & 59.44 & 22.57 & 7.85 \\ 
MC-Net+ \citep{wu2022mutual} (MIA'22)  & & & 70.00 & 55.66 & 16.03 & 3.87  \\
SS-Net \citep{wu2022exploring} (MICCAI'22)& & & 65.51 & 51.09 & 18.13 & 3.44 \\ 
MCCauSSL \citep{miao2023caussl} (ICCV'2023) & & & 72.89 & 58.06 & 14.19 & 4.37 \\
QDC-Net \citep{zhang2024quality} (CBM'2024) & & & 76.62 & 62.70 & 9.25 & 2.49 \\
MLRPL \citep{su2024mutual} (MIA' 2024) & & &  75.93	& 62.12	&9.07	&1.54\\
Ours & & & \textbf{77.36}{\color{red}\textbf{\scriptsize{$\uparrow$}1.43}}&\textbf{63.65}{\color{red}\textbf{\scriptsize{$\uparrow$}0.95}}&\textbf{7.13}{\color{red}\textbf{\scriptsize{$\downarrow$}1.94}}& 1.58 \\ \hline
 UA-MT \citep{yu2019uncertainty} (MICCAI’19) & \multirow{7}{*}{12(20\%)} & \multirow{7}{*}{50(80\%)} 

& 76.10 & 62.62 & 10.84 & 2.43 \\
SASSNet \citep{li2020shape} (MICCAI'20) & & & 76.39 & 63.17 & 11.06 & 1.42 \\ 
DTC \citep{luo2021semi} (AAAI'21) &  &  & 76.27 & 62.82 & 8.70 & 2.20  \\
URPC \citep{luo2021efficient} (MICCAI'21) & & & 80.02 & 67.30 & 8.51 & 1.98  \\ 
MC-Net+ \citep{wu2022mutual} (MIA'22) & & & 79.37 & 66.83 & 8.52 & 1.72   \\
SS-Net \citep{wu2022exploring} (MICCAI'22)& & & 76.20 & 63.00 & 10.65 & 2.68 \\
MCCauSSL \citep{miao2023caussl} (ICCV'2023) & & & 80.92 & 68.26 & 8.11 & 1.53 \\
QDC-Net \citep{zhang2024quality} (CBM'2024) & & & 81.23 & 68.77 & 6.17 & 1.68 \\
 MLRPL \citep{su2024mutual} (MIA' 2024)  & & & 81.53 & 69.35 & 6.81 & 1.33\\
Ours & & & \textbf{83.66}{$\color{red}\textbf{\scriptsize{$\uparrow$}2.13}$}&\textbf{72.14}{\color{red}\textbf{\scriptsize{$\uparrow$}2.79}}&\textbf{4.66}{\color{red}\textbf{\scriptsize{$\downarrow$}2.15}}&\textbf{1.27}{\color{red}\textbf{\scriptsize{$\downarrow$}0.06}}\\ \hline
\end{tabular}}
}
\caption{Comparisons with state-of-the-art semi-supervised segmentation methods on pancreas dataset. Improvements compared with the second-best results are highlighted in \textcolor{red}{Red}.}
\label{tab:pancreas}
\end{table*}

\subsection{Overview of our framework}

The proposed LLM-SegNet as shown in Figure \ref{fig:main} employs a co-training mechanism with two networks. Learning from annotated data is performed using a supervised loss. However, for leveraging the unannotated data, each model utilizes the output of the other model to generate pseudo-labels for its training which facilitates collaborative learning. Both networks share the same model architecture, but to enhance diversity in both models, we employ the CutMix augmentation \citep{ghiasi2021simple}. In our approach, Model $\mathcal{A}$ utilizes the original unlabeled image, while Model $\mathcal{B}$ employs the CutMix technique with an unlabeled image. By having one model see the original image and another see the CutMix image, the models are exposed to different perspectives of the same data. This can help the models learn more diverse representations of the data and potentially discover complementary features that can improve overall performance.


We first apply the CutMix technique to the unlabeled image, resulting in a cut-mixed image. Next, we feed this image into Model $\mathcal{B}$ to generate its unlabeled prediction. Now, to ensure consistency while applying the unsupervised loss between the prediction of the cut-mixed image from Model $\mathcal{B}$ and the pseudo-label from Model $\mathcal{A}$, we also subject the pseudo-label of Model $\mathcal{A}$ to the CutMix technique. This ensures consistency while applying the unsupervised loss.


Let $\alpha$ and $\beta$ denote the model parameters for model $\mathcal{A}$ ($f_\alpha(.)$) and model $\mathcal{B}$ ($f_\beta(.)$), respectively. Our goal is to find the values of ${\alpha}^\ast$ and ${\beta}^\ast$ that result in the lowest overall loss when training from both annotated and unannotated data.


\begin{equation}
   {\alpha}^\ast, {\beta}^\ast= \underset{\begin{subarray}{c}
 {\alpha, \beta}
  \end{subarray}}{\text{arg min }}\{ \mathcal{L}_{s}(X_{ann}; \alpha, \beta)+\mathcal{L}_{u}(X_{unann}; \alpha, \beta) \}
\end{equation}

A batch of input data contains the same number of annotated and unannotated images. Additionally, random noise is added to the input data. This noise is intended to perturb the input data, introducing variability into the model's training process. Adding noise to the input data acts as a regularization technique, enhancing the model's robustness and improving its generalization ability. Finally, these images are sent to each model to get the predictions.

\begin{equation}
\hat{\mathbf{y}}_\alpha = f_\alpha(\mathbf{x + \eta}) \quad \text{and} \quad \hat{\mathbf{y}}_\beta = f_\beta(\mathbf{x + \eta})
\end{equation}

$\hat{y}_\alpha$ and $\hat{y}_\beta$ includes the prediction for annotated and unannotated data. $\eta\sim U(-0.2, 0.2)$ represents random noise sampled from a uniform distribution whose range is between -0.2 and 0.2. The annotated data is leveraged for learning through supervised loss $\mathcal{L}_{s}$. Specifically, we utilize voxel-level dice $({dice})$ and cross-entropy loss $({ce})$ between the labeled prediction and ground truth of the image, as given below:

\begin{equation}
\begin{split}
    \mathcal{L}_{s}(X_{ann}; \alpha, \beta)&=\hspace{-0.35cm}\sum_{(x, y)\in X_{ann}}\hspace{-0.5cm}{dice}(\hat{y}_{\alpha}, y)+{dice}(\hat{y}_{\beta}, y) \\
    &\quad + {ce}(\hat{y}_{\alpha}, y)+{ce}(\hat{y}_{\beta}, y)
\end{split}
\end{equation}


Similarly, to train the model on unlabeled data, we utilize the pseudo-labeling framework to enable each model to learn semantic information from the other one. Given a model prediction $\hat{y}_{\alpha}$, the pseudo label generated by it can be written as $\tilde{y}_{\alpha}$, where $\tilde{y}_{\alpha}$ is defined as follows:

\begin{equation}
\Tilde{y}_{\alpha} =
\begin{cases}
\phi(\hat{y}_{\alpha}) & , \text{ if loss function = MSE} \\
argmax(\hat{y}_{\alpha}) & , \text{ otherwise}
\end{cases}
\end{equation}

\noindent where $\phi$ is the softmax operation and the $argmax$ function determines the class with the highest predicted probability.

We utilize the Unified Segmentation loss ($USL$) (Section \ref{sec:loss}) to ensure that the second model is trained not only on voxels reliably classified with higher probability by the first model but also on those exhibiting lower certainty and vice versa. Our unsupervised loss $\mathcal{L}_{u}$ is formalized as follows: 

\begin{equation}
    \mathcal{L}_{u}(X_{unann}; \alpha, \beta)=\hspace{-0.35cm}\sum_{(x)\in X_{unann}}\hspace{-0.5cm}{USL}(\hat{y}_{\alpha}, \Tilde{y}_{\beta})+{USL}(\hat{y}_{\beta}, \Tilde{y}_{\alpha})
\end{equation}



\subsection{Leveraging task-specific textual descriptors from LLM}

LLMs have proven to be effective in giving medical expertise and guidance. ChatGPT, a big dialog-based LLM, has proven to be highly effective in medical knowledge evaluation. Inspired by this, we leverage an LLM to generate descriptive text about the segmentation task to provide important information, including anatomy and spatial context. Some of the responses are shown in Table \ref{tab:llm}. Specifically, we exploit ChatGPT-4 to generate descriptive text ($t$) about the medical images of interest that can facilitate the segmentation task, as shown in Figure \ref{fig:prompt}.

The generated text ($t$) is forwarded to the BERT \citep{DevlinCLT19} tokenizer ($BERT$) that provides word embeddings corresponding to the input text. We then pass word embeddings to an MLP (multilayer perceptron) to generate text features ($z$) and adjust its size to match that of the image features as follows:

\begin{equation}
    z=MLP(BERT(t))
\end{equation}

To incorporate the knowledge obtained from LLM, we add the text features $z$ with the latent vector obtained at the bottleneck of the encoder module (image features $\mathcal{E}(x)$). Mathematically, let for an input image $x$, the segmentation mask for the input image $x$ is obtained as follows: 

\begin{equation}
    M(x)=\mathcal{D}(\mathcal{E}(x)+ \beta * z)
\end{equation}

Here $\mathcal{E}$ and $\mathcal{D}$ signify encoder and decoder module respectively and $\beta$ is the parameter which is used to balance the image and text features.

\begin{table*}[t] \small 
\centering
\scriptsize
\setlength{\tabcolsep}{0.7mm}{
\resizebox{0.7\linewidth}{!}{
\begin{tabular}{c|cc|llll}
\hline
\multicolumn{1}{c|}{\multirow{2}{*}{Method}} & \multicolumn{2}{c|}{Scans used}  & \multicolumn{4}{c}{Metrics} \\ 
\cline{2-7} \multicolumn{1}{c|}{} & \multicolumn{1}{l}{Labeled} & \multicolumn{1}{l|}{Unlabeled} & 
Dice$\uparrow$ & Jaccard$\uparrow$ & 95HD$\downarrow$ & ASD$\downarrow$ \\ \hline
\multicolumn{1}{c|}{U-Net} &\multicolumn{1}{c}{25(10\%)} &\multicolumn{1}{c|}{0} & 74.43 & 61.86 & 37.11 & 2.79 \\ 
\multicolumn{1}{c|}{U-Net} &\multicolumn{1}{c}{50(20\%)} &\multicolumn{1}{c|}{0} & 80.16 & 71.55 & 22.68 & 3.43 \\ \hline

DAN \citep{zhang2017deep} (MICCAI'17)  & \multirow{9}{*}{25(10\%)} & \multirow{9}{*}{225(90\%)} & 81.71 & 71.43  & 15.15 & 2.32 \\
CPS \citep{chen2021semi} (CVPR'21) &  &  & 82.52 & 72.66 & 13.08 & 2.66 \\
EM \citep{vu2019advent} (CVPR'19) &  &  & 82.27 & 72.15  & 11.98 & 2.42 \\
DTC \citep{luo2021semi} (AAAI'21) &  &  & 81.75 & 71.63  & 15.73 & 2.56 \\
CCT \citep{ouali2020semi} (CVPR'20) &  &  & 82.15 & 71.59  & 16.57 & 2.11 \\
URPC \citep{luo2021efficient} (MICCAI'21) &  &  & 82.59 & 72.11  & 13.88 & 3.72 \\
CPCL \citep{xu2022all} (BHI'22) &  & & 83.36 & 73.23 & 11.74 & 1.99 \\
AC-MT \citep{xu2023ambiguity} (MIA'23) &  &  & 83.77 & 73.96  & 11.37 & 1.93 \\
MLRPL \citep{su2024mutual} (MIA'24) & & &  84.29 & 74.74 & 9.57 & 2.55 \\ 
Ours & & & \textbf{85.95}{\color{red}\textbf{\scriptsize{$\uparrow$}1.66}}&\textbf{76.71}{\color{red}\textbf{\scriptsize{$\uparrow$}1.97}}&\textbf{7.62}{\color{red}\textbf{\scriptsize{$\downarrow$}1.95}}&\textbf{1.74}{\color{red}\textbf{\scriptsize{$\downarrow$}0.19}} \\ \hline
DAN \citep{zhang2017deep} (MICCAI'17) & \multirow{7}{*}{50(20\%)} & \multirow{7}{*}{200(80\%)} 
 & 83.31 & 73.53  & 10.86 & 2.23 \\

CPS \citep{chen2021semi} (CVPR'21) &  &  & 84.01 & 74.02 & 12.16  & 2.18 \\
EM \citep{vu2019advent} (CVPR'19) &  &  & 82.91 & 73.15  & 10.52 & 2.48 \\
DTC \citep{luo2021semi} (AAAI'21) &  &  & 83.43 & 73.56  & 14.77 & 2.34 \\
CCT \citep{ouali2020semi} (CVPR'20) &  &  & 82.53 & 72.36  & 15.87 & 2.21\\
URPC \citep{luo2021efficient} (MICCAI'21) &  &  & 82.93 & 72.57  & 15.93 & 4.19 \\
CPCL \citep{xu2022all} (BHI'22) &  &  & 83.48 & 74.08  & 9.53 & 2.08 \\
AC-MT \citep{xu2023ambiguity} (MIA'23) &  &  & 84.63 & 74.39  & 9.50 & 2.11 \\
MLRPL \citep{su2024mutual} (MIA'24) &  &  & 85.47 & 76.32  & 7.76 & 2.00 \\ 
Ours & & & \textbf{86.69}{$\color{red}\textbf{\scriptsize{$\uparrow$}1.22}$}&\textbf{77.57}{\color{red}\textbf{\scriptsize{$\uparrow$}1.25}}&\textbf{6.83}{\color{red}\textbf{\scriptsize{$\downarrow$}0.93}}&\textbf{1.71}{\color{red}\textbf{\scriptsize{$\downarrow$}0.29}}\\ \hline
\end{tabular}}
}
\caption{Comparisons with state-of-the-art semi-supervised segmentation methods on Brats-2019 dataset for 10\% and 20\% labeled data. 
Improvements compared with the second-best results are highlighted in \textcolor{red}{Red}.}
\label{tab:brats2019}
\end{table*}

\subsection{Unified segmentation loss} \label{sec:loss}


Unsupervised loss plays a key role in deciding the final performance of semi-supervised models. Conventional Negative Log Likelihood (NLL) loss encourages the model to make confident and discriminative predictions by penalizing incorrect classifications more heavily. Whereas, mean squared error (MSE) loss measures the squared difference between predictions and pseudo labels, which tends to be less sensitive to label noise compared to NLL loss. This can be beneficial in scenarios where the pseudo-labels may contain inaccuracies or uncertainties. Inspired by this, we apply NLL loss when our model makes confident predictions about foreground and background, as shown in Equations \ref{eq:CE_f} and \ref{eq:CE_b}. However, when the model is not highly confident about the foreground and background predictions, we apply mean squared error loss, as depicted in Equation \ref{eq:MSE}.  We term this combined loss function a ``Unified Segmentation Loss" (USL). Mathematically, we define our loss function for model $\mathcal{A}$ as  follows:


\begin{equation}
\begin{split}
\text{USL}(\hat{y}_\alpha, \tilde{y}_\beta) &= \mathbbm{I}(\hat{y}'_\beta>t_1) \times \text{NLL}((\hat{y}'_\alpha, \tilde{y}_\beta) \\
&\quad + \mathbbm{I}(\hat{y}'_\beta<t_2) \times \text{NLL}(1- \hat{y}'_\alpha, 1- \tilde{y}_\beta) \\
&\quad + \mathbbm{I}(t_2<\hat{y}'_\beta<t_1) \times \text{MSE}(\hat{y}_\alpha, \tilde{y}_\beta)
\end{split}
\end{equation}


As defined earlier, $\hat{y}_\alpha$ represents the raw prediction from model $\mathcal{A}$, and $\tilde{y}_\beta$ is the pseudo-label from model $\mathcal{B}$. Further, $\hat{y}'_\alpha$ = $\phi(\hat{y}_\alpha)$ and $\hat{y}'_\beta$ = $\phi(\hat{y}_\beta)$ where $\phi$ is the softmax operation and $\hat{y}_\beta$ is the raw prediction from model $\mathcal{B}$.



\begin{equation} \label{eq:CE_f}
\mathbb{I}(\hat{y}'_\beta>t_1)= 
\begin{cases}
\max(\hat{y}'_\beta) & , \text{ if } \max(\hat{y}'_\beta) > t_1 \\
0 & , \text{ otherwise }
\end{cases}
\end{equation}
\begin{equation} \label{eq:CE_b}
\mathbb{I}(\hat{y}'_\beta<t_2)= 
    \begin{cases}
1 - \min(\hat{y}'_\beta) & , \text{ if } \min(\hat{y}'_\beta) < t_2 \\
0 & , \text{ otherwise }
\end{cases}
\end{equation}
\begin{equation} \label{eq:MSE}
\mathbb{I}(t_2<\hat{y}'_\beta<t_1)= 
   \begin{cases}
1 & , \text{ if } t_2 < \min(\hat{y}'_\beta) \text{ and } \max(\hat{y}'_\beta) < t_1 \\
0 & , \text{ otherwise }
\end{cases}
\end{equation}
\noindent

Here, NLL and MSE represent the negative log-likelihood and mean square error loss, respectively. $\mathbb{I}$ is a function, and $t_1$ and $t_2$ are threshold values defined empirically.

\begin{figure*}[t]
    \centering
    \includegraphics[width=0.7\textwidth]{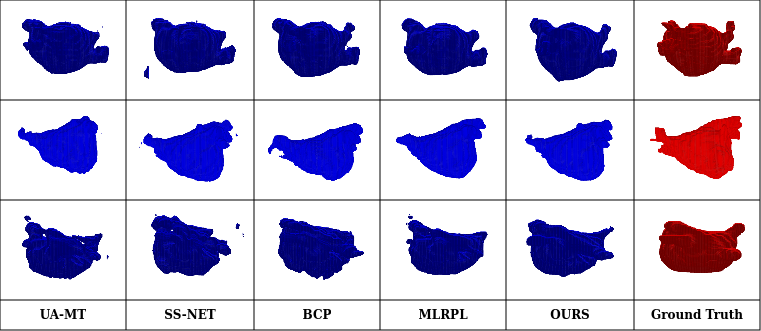}
    \caption{Qualitative results of our method compared to recent state-of-the-art methods on the LA dataset with a 10 \% labeled ratio.}
    \label{fig:LA}
\end{figure*}

\begin{figure*}[t]
    \centering
    \includegraphics[width=0.7\textwidth]{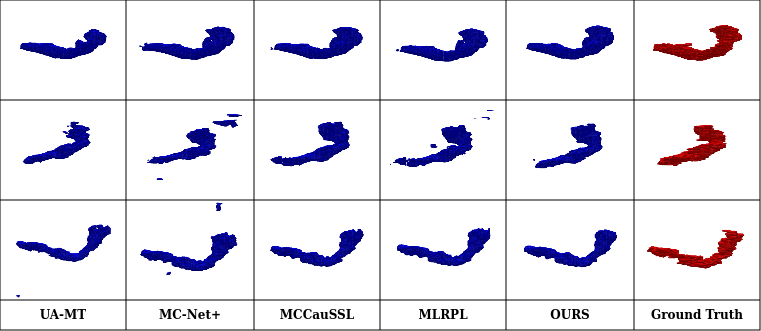}
    \caption{Qualitative results of our method compared to recent state-of-the-art methods on the pancreas dataset with a 20 \% labeled ratio.}
    \label{fig:pan}
\end{figure*}

\section{Experiments}

\subsection{Dataset and experimental protocol}

\textbf{LA dataset:} The left atrium (LA) dataset \citep{xiong2021global} comprises 100 3D gadolinium-enhanced MR images, each possessing an isotropic resolution. Adhering to prior semi-supervised methodologies \citep{li2020shape,bai2023bidirectional}, we partition the 100 scans into 80 samples for training and 20 samples for validation. During the training phase, we randomly extract 3D patches sized 112 $\times$ 112 $\times$ 80 voxels as inputs and employ a sliding window strategy with a stride of 18 $\times$ 18 $\times$ 4 voxels for testing. Notably, for semi-supervised learning on this dataset, in line with experimental protocols used in \citep{wu2022mutual}, only a small fraction of the training data, specifically 5\% and 10\%, is designated as annotated data, while the remaining data is treated as unannotated.

\textbf{Pancreas dataset:} Pancreas-CT dataset \citep{roth2015deeporgan} comprises 82 3D abdominal contrast-enhanced CT scans. These scans maintain a consistent resolution of 512 $\times$ 512 pixels, with thicknesses varying from 1.5 to 2.5 mm. In line with the data partitioning approach described in \citep{wu2022mutual}, we allocate 62 scans for training purposes and evaluate the model's performance on the remaining 20 scans. During training, we randomly extract 3D patches sized 96 $\times$ 96 $\times$ 96 as inputs and employ a sliding window strategy with a stride of 16 $\times$ 16 $\times$ 16 voxels for testing. We utilize 10\% and 20\% of the annotated training data, while the remainder is considered unannotated. 

\textbf{BRATS-2019 dataset:} The BRATS-2019 dataset \citep{bakas2020brats} includes 335 glioma patients' MRI images from different hospitals. The MRI datasets of each patient include four modalities (T1, T1Gd, T2, and T2-FLAIR) with pixel-wise annotations. We employ 250 images for training and 25 for validation and assess performance on the remaining 60 scans based on \citep{su2024mutual}. We use a sliding window technique with a stride of 64 $\times$ 64 $\times$ 64 voxels for testing and randomly extract patches of size 96 $\times$ 96 $\times$ 96 voxels as input during training. We utilize 10\% and 20\% of the annotated training data, while the remainder is considered unannotated. 

\textbf{Implementation details:} We utilize an NVIDIA A5000 GPU with a consistent random seed across all our experiments. To ensure comparability with earlier techniques, we employ V-Net \citep{MilletariNA16} for the pancreas-CT and LA datasets, and U-Net \citep{ronneberger2015u} serves as the backbone for the BraTS-19 dataset. The models are trained using the stochastic gradient descent (SGD) optimizer, with an initial learning rate of $5 \times 10^{-2}$, momentum of 0.9, and weight decay factor of $10^{-4}$ for the LA and BraTS-19 datasets. Similarly, for the pancreatic dataset, the SGD optimizer is employed, starting with a weight decay factor of $10^{-4}$, a momentum of 0.9, and an initial learning rate of $2.5 \times 10^{-2}$.

\textbf{Evaluation metrics:} We choose four complimentary evaluation metrics, which include the Average Surface Distance (ASD), 95\% Hausdorff Distance (95HD), Jaccard coefficient \citep{choi2010survey}, and Dice coefficient \citep{choi2010survey}. While Dice and Jaccard metrics are area-based, 95HD and ASD metrics evaluate boundaries.

\begin{figure*}[t]
    \centering
    \includegraphics[width=0.7\textwidth]{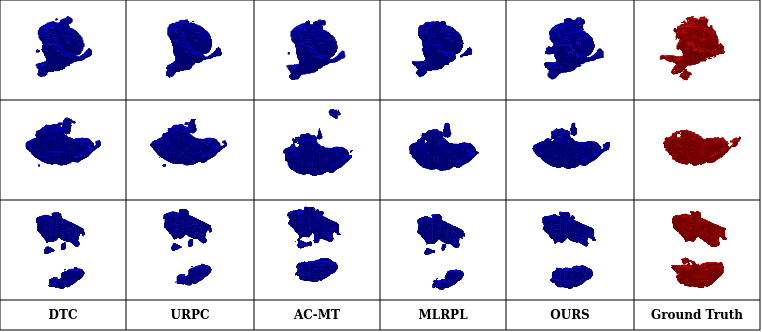}
    \caption{Qualitative results of our method compared to recent state-of-the-art methods on the Brats-2019 dataset with a 10 \% labeled ratio.}
    \label{fig:BRATS}
\end{figure*}

\subsection{Comparison with State-of-the-art techniques for Left atrium dataset}
The proposed LLM-SegNet outperforms state-of-the-art semi-supervised 3D segmentation methods on the Left Atrium (LA) \citep{XiongXHHBZVRMYH21} dataset. Results obtained on the LA dataset with only 5\% annotated samples are provided in Table \ref{tab:LA8} demonstrate that LLM-SegNet outperforms the Dice score by 0.61\%, Jaccard score by 0.81\%, 95HD by 0.28 and ASD by 0.02. Likewise, with 10\% annotated samples, the proposed LLM-SegNet outperforms the Dice score by 1.59\%, Jaccard score by 2.63\%, 95HD by 2.15, and ASD by 0.12. Further, to qualitatively demonstrate the superiority of our method, we provide visualization results comparing it with recent state-of-the-art methods, as shown in Figure \ref{fig:LA}. Our method exhibits better segmentation accuracy compared to the others.


\subsection{Comparison with State-of-the-art techniques for pancreas dataset}

Given the challenging anatomical structure of the pancreas, segmentation of this organ poses difficulties. To demonstrate the effectiveness of our method, we conduct experiments on this dataset. Table \ref{tab:pancreas} presents the results on the Pancreas dataset, with 20\% annotated samples, the proposed LLM-SegNet outperforms the Dice score by 2.13\%, Jaccard score by 2.79\%, 95HD by 2.15, and ASD by 0.06. These results establish the superiority of the proposed LLM-SegNet over the state-of-the-art. Additionally, we offer visualizations to showcase the effectiveness of our approach compared to recent state-of-the-art methods, as shown in Figure \ref{fig:pan}. Our method demonstrates superior segmentation accuracy when compared to others.

\subsection{Comparison with State-of-the-art techniques for Brats-19 dataset}

To further demonstrate the effectiveness of our method, we conduct experiments on the Brats19 dataset \citep{bakas2020brats}. Table \ref{tab:brats2019} presents the results on the Brats19 dataset, with 10\% annotated samples the proposed LLM-SegNet outperforms the Dice score by 1.66\%, Jaccard score by 1.97\%, 95HD by 1.95 and ASD by 0.19. Similarly, with 20\% annotated samples, the proposed LLM-SegNet outperforms the Dice score by 1.22\%, Jaccard score by 1.25\%, 95HD by 0.93, and ASD by 0.29. These results establish the superiority of the proposed LLM-SegNet over the state-of-the-art. Further, to qualitatively demonstrate the superiority of our method, we provide visualization results comparing it with recent state-of-the-art methods, as shown in Figure~\ref{fig:BRATS}. Our method exhibits better segmentation accuracy compared to the others.


\section{Ablation study}

\subsection{Significance of LLM}
We quantify the contribution to improved segmentation performance through leveraging LLM. To be specific, we examine the influence of the beta parameter $\beta$ on the segmentation performance for the pancreas and LA dataset. $\beta$ is the hyperparameter that is used to balance the image and text features. Table \ref{tab:token_para} shows that boosting the hyperparameter $\beta$ significantly improves scores for all evaluation measures in the LA dataset. Similarly, raising $\beta$ from 1 to 2 somewhat improves metrics dice and jaccard but raises the value of metrics 95HD and ASD, therefore we chose $\beta = 1$ for the pancreatic dataset. We chose $\beta = 1$ for the Brats19 dataset as well. The results clearly show that the infusion of information about the medical image segmentation task obtained from the LLM improves the segmentation outcomes.


\begin{table}[t] \small 
\centering
\scriptsize
\setlength{\tabcolsep}{0.4mm}{
\resizebox{0.99\linewidth}{!}{
\begin{tabular}{c|cc|c|llll}
\hline
\multicolumn{1}{c|}{\multirow{2}{*}{Dataset}}  & \multicolumn{2}{c|}{Scans used}  & \multicolumn{1}{c|}{\multirow{2}{*}{$\beta$}} & \multicolumn{4}{c}{Metrics} \\ 
\cline{2-3} \cline{5-8} &\multicolumn{1}{l}{Labeled} & \multicolumn{1}{l|}{Unlabeled} &  & 
Dice$\uparrow$ & Jaccard$\uparrow$ & 95HD$\downarrow$ & ASD$\downarrow$ \\ \hline
\multirow{3}{*}{Pancreas}  & \multirow{3}{*}{12(20\%)}& \multirow{3}{*}{50(80\%)}  & 0.1 &  83.11 & 71.32 & 9.37 & 2.48  \\
 &  & & 1   &  83.66 & 72.14 & \textbf{4.66} & \textbf{1.27} \\ 
 &  & & 2   &  \textbf{83.70} & \textbf{72.16} & 7.54 & 2.18 \\  \hline

\multirow{3}{*}{LA} & \multirow{3}{*}{8(10\%)}& \multirow{3}{*}{72(90\%)}  & 0.001 & 90.49 & 82.73 & 5.93 & 1.94  \\
 &  & & 0.01   & 90.88 & 83.37 & 5.03 & 1.72 \\ 
 &  & & 0.1   &  \textbf{91.45} & \textbf{84.31} & \textbf{4.66} & \textbf{1.62} \\  \hline

 \end{tabular}}
}
\caption{Ablation experiment on the value of hyperparameter $\beta$ for pancreas and LA datasets.}
\label{tab:token_para}
\end{table}

\begin{table}[t]
\centering
\resizebox{0.48\textwidth}{!}{%
\begin{tabular}{lcccc}
\hline
\textbf{Method} & \textbf{Dice (\%)$\uparrow$} & \textbf{Jaccard (\%)$\uparrow$} & \textbf{95HD (voxel)$\downarrow$} & \textbf{ASD (voxel)$\downarrow$} \\ \hline
CE & 83.08 & 71.34 & 5.22 & 1.64 \\ 
DICE & 82.30 & 70.23 & 5.89 & 1.71 \\ 
CE + DICE & 83.66 & 72.14 & 4.66 & 1.27 \\ \hline

\end{tabular}
}
\caption{Comparison of different supervised losses on pancreas dataset with 20\% labeled ratio.}
\label{tab:suploss}
\end{table}

\begin{figure*}[t]
    \centering
    \includegraphics[width=0.85\textwidth]{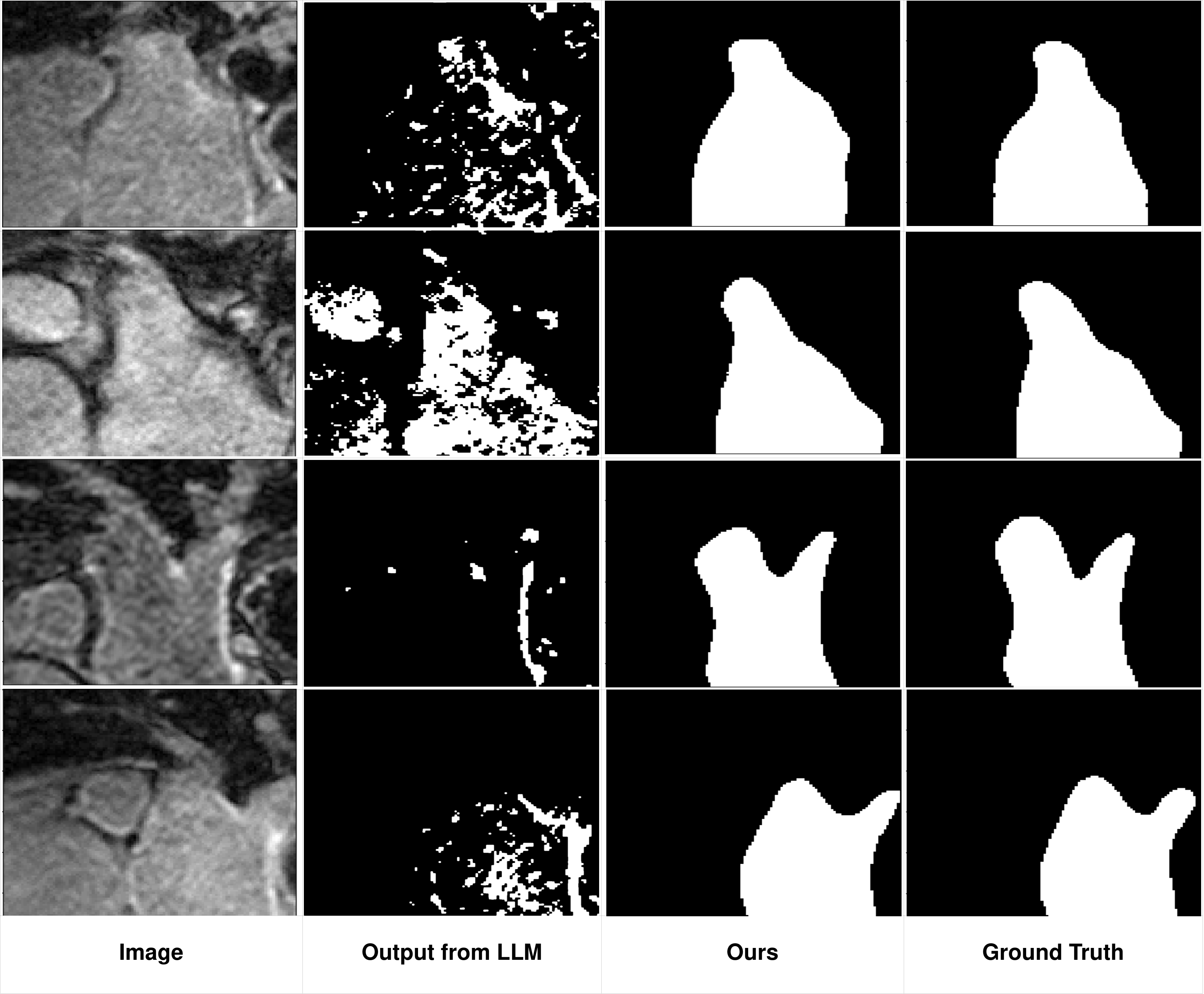}
    \caption{Comparative results of using ChatGPT-4 for segmentation purposes are presented alongside our method. To segment the organ, we provide the following prompt to ChatGPT-4: ``Generate a binary segmentation mask for this image. The image is a 2D slice extracted from a 3D image of the heart. It contains the image of the left atrium, one of the four chambers of the heart. The segmentation mask should have the same dimensions as the input image and should highlight the region corresponding to the left atrium. Pay attention to any distinctive features or landmarks that indicate the presence of the left atrium. Please consider any potential noise or artifacts in the image and aim to produce a clean segmentation."}
    \label{fig:llmcompare}
\end{figure*}

\begin{table}[t]
\centering
\resizebox{0.48\textwidth}{!}{%
\begin{tabular}{lcccc}
\hline
\textbf{Method} & \textbf{Dice (\%)$\uparrow$} & \textbf{Jaccard (\%)$\uparrow$} & \textbf{95HD (voxel)$\downarrow$} & \textbf{ASD (voxel)$\downarrow$} \\ \hline
MSE & 80.16 & 67.32 & 8.93 & 2.31 \\
NLL & 83.33 & 71.62 & 10.26 & 3.59 \\
USL & 83.66 & 72.14 & 4.66 & 1.27 \\
\hline
\end{tabular}%
}

\caption{Comparison of different unsupervised losses on pancreas dataset (setting 12:50).}
\label{tab:unsuploss}
\end{table}

\subsection{Textual response vs. visual response from LLM} \label{ablation_2}

In this section, we demonstrate why we chose to utilize textual responses from LLM instead of directly relying on their segmentation outputs. To illustrate this, we compare our method with the output obtained from ChatGPT-4, as depicted in Figure \ref{fig:llmcompare}. As seen, ChatGPT-4 struggles to accurately segment the LA organ from the provided 2D slices, while our approach demonstrates a superior overlap rate with the ground truth. This highlights the limitations of solely depending on LLM for medical image segmentation.



\subsection{Significance of supervised losses}
The supervised loss $\mathcal{L}_{s}$ comprises the sum of cross entropy and dice loss. We study the contribution of each component of the supervised loss term toward improved segmentation performance. Results reported in Table \ref{tab:suploss} demonstrate that cross-entropy loss is more useful versus dice loss as it renders better scores on all the evaluation metrics. Furthermore, as expected, incorporating both cross-entropy and dice loss renders the best segmentation performance, as quantified through all the evaluation metrics.

\subsection{Significance of Unified segmentation loss}
The unsupervised loss $\mathcal{L}_{u}$ used to train the model leverages our proposed unified segmentation loss ($USL$) that comprises negative log-likelihood ($nll$) and mean square error ($mse$) loss. We study the contribution of each component of the unsupervised loss term toward improved segmentation performance. Results reported in Table \ref{tab:unsuploss} signify that negative log-likelihood loss is more useful compared to mean square error loss as it renders better scores on all the evaluation metrics. USL demonstrates superior performance across all four metrics when compared to NLL. The improvements in Dice and Jaccard scores, though small, suggest a slightly better overall segmentation accuracy. The significant reductions in both 95HD and ASD indicate that USL produces more precise and consistent segmentations, with fewer large errors and a closer fit to the ground truth surfaces. 

The higher ASD and 95HD values for NLL compared to USL suggest that NLL is less effective at producing accurate and consistent boundary predictions. This can be attributed to NLL's property of making confident and discriminative predictions by heavily penalizing incorrect classifications. However, in areas with uncertainties, particularly at the edge regions in medical images, NLL shows less performance, as evidenced by the higher ASD and 95HD values. Conversely, our own loss function, which uses MSE loss for areas where the model is less confident, yields better performance in terms of ASD and 95HD. This is because MSE is less punitive in uncertain regions, allowing the model to produce smoother and more accurate boundary predictions. This approach results in improved performance in these critical areas, leading to lower ASD and 95HD values.


\section{Conclusion \label{conclusion}}

Semi-supervised learning facilitates learning from limited annotated data and abundant unlabeled data. In this paper, we present LLM-SegNet, a novel semi-supervised learning approach that harnesses the power of large language models to infuse task-specific knowledge into the co-training framework, thereby enhancing learning from unannotated data. This knowledge assists the model in effectively understanding the characteristics of the region of interest (ROI), ultimately facilitating more efficient segmentation. Furthermore, we propose a unified segmentation loss function. This loss function not only prioritizes regions where the model is confident but also effectively handles areas where it lacks confidence in foreground and background predictions. Results on three publicly available 3D medical segmentation datasets confirm the efficacy of the proposed approach.

\bibliographystyle{elsarticle-num}
\bibliography{refs}

\end{document}